\documentclass[tightenlines,aps,floatfix,preprint]{revtex4}
\usepackage{psfig,float,axodraw}

\newcommand{\ba}{\begin{eqnarray}}
\newcommand{\ea}{\end{eqnarray}}

\newcommand{\ep}{\epsilon}

\newcommand{\be}{\begin{equation}}
\newcommand{\ee}{\end{equation}}

\begin{document}

\title{
\hfill{\normalsize\vbox{%
\hbox{\rm SLAC-PUB-10252}
\hbox{\rm UH-511-1039-03}
}}
\vspace*{0.5cm}

A new method for real radiation at NNLO
}

\author{
Charalampos Anastasiou\thanks{e-mail:babis@slac.stanford.edu}} 
\affiliation{
          Stanford Linear Accelerator Center,\\ 
          Stanford University, Stanford, CA 94309, U.S.A.}
\author{Kirill Melnikov
        \thanks{e-mail: kirill@phys.hawaii.edu}}
\affiliation{Department of Physics and Astronomy,
          University of Hawaii,\\ 2505 Correa Rd., Honolulu, Hawaii 96822}  
\author{Frank Petriello\thanks{frankp@pha.jhu.edu}}
\affiliation{
Department of Physics, Johns Hopkins University, \\
3400 North Charles St., Baltimore, MD 21218
}

\begin{abstract}

We propose a new method of computing real emission contributions to hard 
QCD processes.  Our approach uses sector decomposition of the exclusive 
final-state phase space to enable extraction of all singularities of the 
real emission matrix elements before integration over any kinematic 
variable.  
The exact kinematics of the 
real emission process are preserved in all regions of phase space.  
Traditional approaches to extracting singularities from real emission matrix 
elements, such as phase space slicing and dipole subtraction, require both 
the determination of counterterms for double real emission
amplitudes in singular kinematic limits and the integration of these contributions
analytically to cancel the resulting singularities against virtual corrections.  
Our method addresses both of these issues.  The implementation of constraints on 
the final-state phase space, including various jet algorithms, is simple using 
our approach.  We illustrate our method using $e^+ e^- \rightarrow$ jets at 
${\cal O}(\alpha_{S}^2)$ as an example.

\end{abstract}

\maketitle

\section{Introduction}

The future high energy collider physics experimental program will measure 
phenomenologically interesting quantities with an unprecedented precision.  
To fully utilize these results, accurate theoretical predictions are required.  
In particular, the large value of the strong coupling constant $\alpha_S$ 
implies that perturbative QCD corrections through next-to-next-to leading order 
(NNLO) in $\alpha_S$ are needed.  The calculation of NNLO QCD corrections has 
advanced rapidly in the past few years.  The progress has resulted primarily 
because of the realization that the computation of two-loop virtual corrections 
can be algorithmically structured and automated.  These advances culminated in 
the evaluation of two-loop virtual corrections for $1 \rightarrow 3$  and  all massless 
$2 \rightarrow 2$ scattering processes in perturbative QCD~\cite{glover}.

Unfortunately, these calculations have not yet produced improved theoretical 
predictions for many observables.  The computation of infrared-safe quantities 
requires two additional ingredients: the real-virtual contributions, which denote 
the one-loop corrections to processes with one additional parton in the final state, 
and the real-real contributions, which denote tree-level processes with two 
additional partons in the final state.  The two components of the real-virtual 
contributions, one-loop virtual corrections and single emission amplitudes, have 
both been well studied.  However, the double emission corrections 
required for the real-real contributions are relatively unknown; the first steps 
towards understanding them have been taken only 
recently~\cite{Kosower:2003cz,Weinzierl:2003fx}.

The current state-of-the-art can be 
illustrated using $e^+e^- \rightarrow$ 2 jets 
as an example.  Unresolved double real emission corrections appear for the 
first time at NNLO.  The 2 jet cross section is currently computed at NNLO 
by taking the difference of the ${\cal O}(\alpha_{S}^{2})$ $e^+e^- \rightarrow$ 
hadrons cross section and the $e^+e^- \rightarrow$ 3 and 4 jet results 
at NLO and LO, respectively.  Only the total 2 jet cross section can be 
derived using this technique.  All information concerning the invariant mass 
and angular distributions of the jets is lost.

The inability to compute the 2 jet cross section directly at NNLO arises from 
the  
poor understanding of the singular structure of the double real emission 
corrections.  Infrared and 
collinear singularities cancel
between virtual and real corrections only after integration over certain kinematic variables 
makes the $1/\ep$ poles in the real emission contributions explicit.  
However, since a primary goal of computing higher 
order QCD corrections to scattering processes is to produce Monte Carlo event 
generators that correctly describe the kinematics of each partonic 
event, only a restricted region of the final-state phase space 
can be integrated over.  Only near the edges of the available phase space, 
where two or more partons become 
degenerate and combine to form a single jet, can the integration be performed without 
changing the kinematics of the final state.  
All singularities occur in these limits, and they can in principle be 
extracted and cancelled against those arising from virtual 
corrections.  Unfortunately, these singularities overlap;
 this severly complicates their extraction.

The existing approaches to computing double real emission corrections extend 
methods used to handle single real emission amplitudes.  There are 
two standard 
techniques for extracting single real emission singularities: phase space 
slicing~\cite{Ellis:1980wv,Giele:1991vf} and dipole 
subtraction~\cite{Fabricius:1980fg,Gutbrod:1983qa,Catani:1996vz}.  
Extending these approaches to double real emission corrections requires 
two non-trivial steps: a determination of the simplified matrix elements 
that approximate the complete double real emission amplitudes in singular 
kinematic regions, and an integration of these matrix elements over the 
unresolved regions of the multi-particle phase space.  The difference of 
the exact and approximate matrix elements is by construction finite, and 
can be integrated numerically.  The integration of the approximate matrix 
elements over the phase space boundaries produces the required 
$1 / \ep$ poles that cancel against the virtual corrections.  
Both steps must be completed to obtain an NNLO prediction.  
Although some progress has recently been made~\cite{Kosower:2002su, 
Kosower:2003cz,Weinzierl:2003fx,Weinzierl:2003ra}, a functional 
method for calculating NNLO real emission corrections has not yet been demonstrated; a substantial 
effort is still required to obtain phenomenological results.

We present here a new approach to this problem.  We 
illustrate our technique by considering the extraction of singularities from 
$1 \rightarrow 4$ processes, where the final state particles are massless.  Our approach  is based upon a few  ideas.  
We first derive a factorized parametrization of the 
four particle phase space following a simple procedure.  
We, then,  use sector 
decomposition~\cite{Binoth:2000ps,Heinrich:2002rc,Binoth:2003ak} 
to separate 
the overlapping divergences which appear in the double real emission matrix elements.  
After this separation is performed, the phase space singularities can be extracted 
using standard expansions in terms of plus distributions.  
The processes of finding the required sectors and extracting 
the singularities are completely automated.  The resulting matrix elements are finite and 
fully differential, and can be used to create NNLO Monte Carlo event generators.  We 
discuss in some detail the example of $e^+e^- \rightarrow$ hadrons at 
${\cal O}(\alpha_{S}^{2})$, which includes the 2 jet cross section at NNLO, the three 
jet cross section at NLO, and the 4 jet cross section at LO.

The paper is organized as follows.  In Section II we introduce our method by considering 
$e^+e^- \to 2~{\rm jets}$ at NLO and $e^+e^- \to 3~{\rm jets}$ at tree level.  
We begin our discussion of the 2 jet cross section at NNLO in Section III by 
describing our parameterization of the four-particle phase space.  We also explain 
how we use sector decomposition to separate the overlapping singularities that appear 
in the matrix elements.  In Section IV we apply our technique to the two most 
difficult interferences that appear in the double real emission contributions.  After 
demonstrating that our method is powerful enough to handle the most complicated 
scenario, we apply it to a simple but realistic example in Section V: 
the $N_f$ dependent contribution 
to $e^+e^- \rightarrow$ hadrons at ${\cal O}(\alpha_{S}^{2})$.  This process contributes 
to the 2 jet cross section at NNLO, the 3 jet cross section at NLO, and the 4 jet 
cross section at LO.  Finally, we present our conclusions and discuss future prospects 
in Section VI.

\section{The NLO example}

We begin by considering the ${\cal O}(\alpha_s)$ contribution 
to $e^+e^- \to {\rm hadrons}$, which contains both the NLO correction 
to $e^+e^- \to 2~{\rm jets}$ and the LO contribution 
to $e^+e^- \to 3~{\rm jets}$.  Although many of the complexities of the NNLO 
case are absent in this calculation, it illustrates several important features 
of our method.  At the partonic level, we must compute the one-loop 
virtual corrections to $e^+e^- \to q \bar q$ and the real emission 
process $e^+e^- \to q \bar q g$.

We consider first the real emission correction $e^+e^- \to q \bar q g$.
The kinematics of the final state is fully described by 
the invariant masses $s_{q \bar q}, s_{q g},$ and $s_{\bar q g}$, which satisfy the constraint
\begin{equation}
s_{q \bar q} + s_{q g} + s_{\bar q g} = s.
\end{equation}
Here, $s$ is  the center of mass energy squared of the colliding electron and positron.  
The three  particle phase-space can be written as
\begin{eqnarray}
&& \int [{\rm d}q][{\rm d}\bar q] [{\rm d} g]
~(2\pi)^d \delta^{(d)}(p_1+p_2 - q - \bar q - g)
=
\nonumber \\ && =
\frac{1}{(4\pi)^{d/2}}
\frac{{\cal R}_2~s^{1-2\ep}}{\Gamma(1-\ep)}
\int \limits_{0}^{1} 
{\rm d}\lambda_1 {\rm d}\lambda_2 \lambda_1^{-\ep} (1-\lambda_1)^{-\ep}
\lambda_2^{-\ep} (1-\lambda_2)^{1-2\ep},
\label{phase3}
\end{eqnarray}
where $d=4-2\ep$,  $[{\rm d}k] = d^{d-1}k/(2\pi)^{d-1}k_0$,
${\cal R}_2$ is the integrated phase-space of the two massless particles,
\begin{equation}
{\cal R}_2 = \frac{1}{(2\pi)^{d-2}}\frac{\Omega_{d-1}}{2^{d-1}},
\end{equation}
and $\Omega_{d}$ is the solid angle in $d$ dimensions,
\begin{equation}
\Omega_d = \frac{2\pi^{\frac{d}{2}}}{\Gamma( \frac{d}{2})}.
\end{equation}
The invariant 
masses have the following expressions in terms of $\lambda_1$ and $\lambda_2$:
\begin{equation}
\label{massesNLO}
s_{q\bar q} = s(1-\lambda_2)(1-\lambda_1),~~~ 
s_{q g} = s(1-\lambda_2) \lambda_1,~~~~s_{\bar q g} = s\lambda_2.
\end{equation}
In what follows we set $s=1$ for simplicity, and restore the correct
dimensions in final results.

The matrix element for the $\gamma^* \to q \bar q g$ process is given
by two diagrams. Upon squaring these and 
 using the expressions for the invariant masses 
given in Eq.~\ref{massesNLO}, we derive~\cite{Ellis:1980wv}
\begin{equation}
|{\cal M}|^2 = \frac{32(1-\ep)}{\lambda_1 \lambda_2
(1-\lambda_2)}
\left (
2 (1-\lambda_1)(1-\lambda_2)+\lambda_2^2
+\lambda_1^2(1-\lambda_2)^2
-\ep \left [ \lambda_1+\lambda_2-\lambda_1\lambda_2   \right ]^2
\right ).
\end{equation}
After substituting the expression for the matrix element squared 
into the three-particle phase-space, we arrive at the expression
\begin{equation}
\int_{0}^{1}{\rm d}\lambda_1{\rm d}\lambda_2 \, 
\frac{{\rm d}^2\sigma_{R}}{{\rm d}\lambda_1 {\rm d} \lambda_2} =  
\frac{1}{(4\pi)^{d/2}}
\frac{{\cal R}_2}{\Gamma(1-\ep)} \int \limits_{0}^{1} 
{\rm d}\lambda_1 {\rm d}\lambda_2 (\lambda_1 \lambda_2)^{-\ep-1} 
(1-\lambda_1)^{-\ep}
(1-\lambda_2)^{-2\ep} g(\lambda_1, \lambda_2),
\label{NLOcr1}
\end{equation}
where $g(\lambda_1,\lambda_2)$ is a non-singular function of the $\lambda_i$.  
The phase space singularities in the above expression can be extracted before 
integration by using the standard decomposition in terms of plus distributions:
\begin{equation}
\lambda^{-1+\ep} = \frac{1}{\ep}\delta(\lambda) + \sum_{n=0}^{\infty} \frac{\ep^{n}}{n!}
\left[  \frac{\ln^{n}(\lambda)}{\lambda} \right ]_+ \, , 
\label{decomp}
\end{equation}
where a plus distribution is defined via
\begin{equation}
\int_0^1 d\lambda 
\left[  \frac{\ln^{n}(\lambda)}{\lambda} \right ]_+ f(\lambda)
= \int_0^1 d\lambda \ln^{n}(\lambda) 
\left[ \frac{f(\lambda) -f(0)}{\lambda} \right].
\end{equation}
Substituting this decomposition into Eq.~\ref{NLOcr1}, 
we derive the following expression for the real emission cross section:
\begin{eqnarray}
\label{NLOreal}
\frac{{\rm d}^2\sigma_{R}}{{\rm d}\lambda_1 {\rm d} \lambda_2}
&=& \frac{64\pi\alpha_s}{3} \sigma_0\frac{\Gamma(1+\ep)}{(4\pi)^{d/2}}\left\{
\frac{\, \delta(\lambda_1)\delta(\lambda_2)}{\ep^2}
+\frac{1}{\ep}\left[-\frac{\delta(\lambda_1)}{[\lambda_2]_+}
-\frac{\delta(\lambda_2)}{[\lambda_1]_+}
+\left(1-\frac{\lambda_1}{2}\right)\delta(\lambda_2) \right. \right. \nonumber \\ & & \hspace{-1.0cm}
+\left(1-\frac{\lambda_2}{2}\right)\delta(\lambda_1)
-\delta(\lambda_1)\delta(\lambda_2)\, \bigg] 
+\left(\lambda_1-1+\frac{1}{2}\frac{(2-2\lambda_1+\lambda_{1}^{2})\,{\rm ln}(1-\lambda_1)}{\lambda_1}
\right. \nonumber \\ & & \hspace{-1.0cm}
-\left(1-\frac{\lambda_1}{2}\right){\rm ln}(\lambda_1) +\left[\frac{1}{\lambda_1}\right]_{+} 
+\left[\frac{{\rm ln}(\lambda_1)}{\lambda_1}\right]_{+} \, \bigg)\, \delta(\lambda_2)  
+\left(\lambda_2-1+\frac{(2-2\lambda_2+\lambda_{2}^{2})\,{\rm ln}(1-\lambda_2)}{\lambda_2}
\right. \nonumber \\ & & \hspace{-1.0cm}
-\left(1-\frac{\lambda_2}{2}\right){\rm ln}(\lambda_2) +\left[\frac{1}{\lambda_2}\right]_{+} 
+\left[\frac{{\rm ln}(\lambda_2)}{\lambda_2}\right]_{+}\,  \bigg)\,\delta(\lambda_1)  
-\left(1-\frac{\lambda_1}{2}\right)\left[\frac{1}{\lambda_2}\right]_{+}  \nonumber \\ & & \hspace{-1.0cm}
-\left(1-\frac{\lambda_2}{2}\right)\left[\frac{1}{\lambda_1}\right]_{+} +\left[\frac{1}{\lambda_1}\right]_{+}
\left[\frac{1}{\lambda_2}\right]_{+} -\frac{\pi^2}{6}\delta(\lambda_1)\delta(\lambda_2)+1-\lambda_1
\left(1-\frac{\lambda_2}{2}\right)  \,  \bigg\} .
\end{eqnarray}
$\sigma_0$ is the tree level cross section for 
$e^+e^- \rightarrow q {\bar q}$: 
$\sigma_0=4\pi\alpha_{{\rm EM}} Q_{q}^{2}/s$.
For the calculation of the NLO corrections to the 2 jet cross section we also require the 
virtual corrections to the $e^+e^- \rightarrow q{\bar q}$ process; 
we find~\cite{Ellis:1980wv}
\begin{equation}
\label{NLOvirt}
\frac{{\rm d}\sigma_{V}}{{\rm d}\lambda_1 {\rm d}\lambda_2}=\frac{64\pi\alpha_s}{3} 
\sigma_0\frac{\Gamma(1+\ep)}{(4\pi)^{d/2}}
\left\{-\frac{1}{\ep^2}-\frac{1}{2\ep}+\frac{2\pi^2}{3}-\frac{5}{2} 
\right\}\delta(\lambda_1)\delta(\lambda_2).
\end{equation}

We now discuss the calculation of the $n$ jet cross section; here, $n$ equals either 2 or 3.  
We introduce the jet function
\begin{equation}
F^{(n)}_{J}(s_{q{\bar q}},s_{qg},s_{{\bar q}g}) 
= F^{(n)}_{J}((1-\lambda_1)(1-\lambda_2),
\lambda_1(1-\lambda_2),\lambda_2).
\end{equation}
The $n$ jet cross section becomes
\begin{equation}
\sigma_{J}^{(n)} = \int_{0}^{1} {\rm d}\lambda_1 {\rm d}\lambda_2 \, F^{(n)}_{J}(s_{ij}) \left\{ \sigma_o 
\delta(\lambda_1)\delta(\lambda_2) 
+ \frac{{\rm d}\sigma_V}{{\rm d}\lambda_1 {\rm d}\lambda_2} + \frac{{\rm d}\sigma_R}{{\rm d}\lambda_1 
{\rm d}\lambda_2} \right\}.
\end{equation}
It is clear from the expressions in Eqs.~\ref{NLOreal} and~\ref{NLOvirt} 
that the $1/\ep^2$ poles cancel when 
$\sigma_V$ and $\sigma_R$ are combined.  The $1/\ep$ poles 
in $\sigma_R$ require that either $\lambda_1$ or $\lambda_2$ vanish.  The jet 
function becomes either $F^{(n)}_{J}(s_{q{\bar q}},s_{qg},0)$ 
or $F^{(n)}_{J}(s_{q{\bar q}},0,s_{{\bar q}g})$ 
in these cases, {\it i.e.}, a 2 jet configuration is always obtained.  
The $1/\ep$ poles of both $\sigma_V$ and $\sigma_R$ occur in the 
2 jet cross section; they cancel after 
integrating over $\lambda_1$ and $\lambda_2$, as required for 
infrared safe observables.  
Dropping the poles in $\ep$, we can write the $n$ jet cross section as an integral over the finite component of the 
partonic cross sections:
\begin{equation}
\sigma_{J}^{(n)}=\int_{0}^{1} {\rm d}\lambda_1 {\rm d}\lambda_2 \, F^{(n)}_{J}(s_{ij})\, \sigma_{finite}.
\end{equation}
It is straightforward to check that 
this correctly reproduces known 2 and 3 jet cross sections for 
standard jet functions~\cite{book}.

Several important aspects of this result generalize immediately to NNLO calculations.  We were able to extract the 
singularities in $\ep$ 
without performing any integrations.  The cancellation of the poles in 
$\ep$ can be 
checked numerically, and these terms can then be discarded.  We note that by casting the subtraction operations needed for 
extracting the $\ep$ poles in terms of plus distributions, 
we have gained the flexibility to combine our result with any jet function 
and  with any constraint on an infrared safe differential quantity.  This greatly simplifies the calculation of NNLO 
cross sections. 

\section{Four particle phase-space: parameterization and 
sector decomposition}

We present here our parameterization of the four particle phase space.  After deriving 
the relevant formulae, we discuss the complications that arise when we attempt to 
extract phase space singularities using the method discussed in the 
previous Section.  
We show how sector decomposition of the phase space solves these problems, and 
apply the technique to a few examples.

We begin with the following expression for the four particle phase space:
\begin{equation}
\label{fourpp1}
{\cal I}_4 = \int [{\rm d}p_1][{\rm d}\bar p_2] [{\rm d} p_3] [{\rm d} p_4]
(2\pi)^d \delta^{(d)}(p-p_1-p_2-p_3-p_4).
\end{equation}
This phase space is described by five independent invariant masses.  A convenient set is 
$\left\{s_{134},s_{234},s_{34},s_{13},s_{24}\right\}$, where $s_{ij}=\left(p_i+p_j\right)^2$ 
and $s_{ijk}=\left(p_i+p_j+p_k\right)^2$.  We split the above integral into three sub-integrals:
\begin{equation}
\label{fourpp2}
{\cal I}_4 =\left(\frac{1}{2\pi}\right)^{3d-4} \int {\rm d}s_{234}{\rm d}s_{34}{\rm d}s_{134}{\rm d}
s_{23}{\rm d}s_{13} I_1I_2I_3,
\end{equation}
where
\begin{equation}
\label{I1}
I_1=\int {\rm d}^d p_1{\rm d}^d Q_{234}\delta^{(d)}\left(p-p_1-Q_{234}\right)\delta\left(p_{1}^2\right)
	\delta\left(Q_{234}^2-s_{234}\right),
\end{equation}
\begin{equation}
\label{I2}
I_2 = \int {\rm d}^d p_2{\rm d}^d Q_{34}\delta^{(d)}\left(Q_{234}-p_2-Q_{34}\right)\delta\left(p_{2}^2\right)
	\delta\left(Q_{34}^2-s_{34}\right),
\end{equation}
and
\begin{equation}
\label{I3}
I_3 = \int {\rm d}^d p_3{\rm d}^d p_4\delta^{(d)}\left(Q_{34}-p_3-p_4\right)\delta\left(p_{3}^2\right)
	\delta\left(p_{4}^2\right).
\end{equation}
We now constrain the integrations over the $s_{ij}$ by introducing the following delta functions 
into ${\cal I}_4$:
\begin{equation}
\label{constaints}
\delta\left(s_{234}-Q_{234}^2\right)\delta\left(s_{34}-Q_{34}^2\right)\delta\left(s_{134}-Q_{134}^2\right)
	\delta\left(s_{23}-2p_2\cdot p_3\right)\delta\left(s_{13}-2p_1\cdot p_3\right).
\end{equation}

To derive representations of these integrals from which the phase space singularities can be conveniently extracted, 
we bring the limits of integration for each integral from 0 to 1 using transformations of the form 
$s_{ij}=\lambda_i\left(s_{ij}^{+}-s_{ij}^{-}\right)+s_{ij}^{-}$, where $s_{ij}^{\pm}$ denote the maximum and minimum 
values of the corresponding invariant masses.  Using the delta functions to simplify the integrations, performing the 
transformation to the variables $\lambda_i$, and including the Jacobian 
$|\partial s_{ij} / \partial \lambda_{k}|$, we arrive at 
\begin{eqnarray}
\label{I4rep1}
{\cal I}_{4}&=& {\cal N}_{4}\int_{0}^{1} {\rm d}\lambda_1{\rm d}\lambda_2{\rm d}\lambda_3{\rm d}\lambda_4{\rm d}\lambda_5 \,
	\delta(\lambda_1-\lambda_{1}^{'})\delta(\lambda_2-\lambda_{2}^{'})\delta(\lambda_3-\lambda_{3}^{'})
	\delta(\lambda_4-\lambda_{4}^{'})\delta(\lambda_5-\lambda_{5}^{'})  \nonumber \\ & & \times
	\left[\lambda_1\left(1-\lambda_1\right)\left(1-\lambda_2\right)\right]^{1-2\ep} 
	\left[\lambda_2\lambda_3
\left(1-\lambda_3\right)\lambda_4\left(1-\lambda_4\right)\right]^{-\ep}
	\left[\lambda_5\left(1-\lambda_5\right)\right]^{-1/2-\ep}.
\end{eqnarray}
We have extracted the overall normalization
\begin{equation}
\label{RRnorm}
{\cal N}_{4}={\cal R}_{2}\left[\frac{\Gamma(1+\ep)}{\left(4\pi\right)^{d/2}}\right]^{2}
\left[\frac{\Omega_{d-1}}{2^{d-1}}\right]^{2}
	\frac{\left(4\pi\right)^d}{\left(2\pi\right)^{2d-2}}
\frac{\Gamma^{2}(2-2\ep)\Gamma(1-2\ep)}{\Gamma^{2}(1+\ep)\Gamma^{4}(1-\ep)
	\Gamma^{2}(1/2-\ep)}.
\end{equation}
The invariant masses have the following expressions in terms of the $\lambda_{i}$ (with $s=1$):
\begin{eqnarray}
\label{invmasses}
s_{234}&=&\lambda_1, \nonumber \\
s_{34}&=&\lambda_1\lambda_2, \nonumber \\
s_{23}&=&\lambda_1\left(1-\lambda_2\right)\lambda_4, \nonumber \\ 
s_{134}&=&\lambda_2+\lambda_3\left(1-\lambda_1\right)\left(1-\lambda_2\right), \nonumber \\
s_{13}&=&\lambda_5\left(s_{13}^{+}-s_{13}^{-}\right)+s_{13}^{-} , 
\end{eqnarray}
with
\begin{equation}
s_{13}^{\pm}=\left(1-\lambda_1\right)\left[\lambda_3\lambda_4+\lambda_2\left(1-\lambda_3\right)\left(1-\lambda_4\right)
	\pm 2\sqrt{\lambda_2\lambda_3\left(1-\lambda_3\right)\lambda_4\left(1-\lambda_4\right)}\right].
\label{s13lims}
\end{equation}

Difficulties arise when we substitute the matrix elements into the four particle phase space 
and attempt to expand the 
expression using Eq.~\ref{decomp}.  
We will discuss in Section V the $N_f$ contributions to 
$e^+e^- \rightarrow$ 2 jets; the matrix elements for the double real 
emission contribution to this 
process contain denominators of the form $1/s_{34}s_{234}s_{134}$.  
Using Eq.~\ref{invmasses}, this 
becomes
\begin{equation}
\label{denomexam}
\frac{1}{s_{34}s_{234}s_{134}} = \frac{1}{\lambda_{1}^{2}\lambda_{2}\left[\lambda_2+\lambda_3
	(1-\lambda_1)(1-\lambda_2)\right]}.
\end{equation}
The third term in this denominator is singular when e.g. 
both $\lambda_2,\lambda_3 \rightarrow 0$, but not when 
only one does.  If we combine the denominator with the integration measure 
in Eq.~\ref{I4rep1}, and 
attempt to naively expand $\lambda_{2}^{-1-\ep} \rightarrow -\delta(\lambda_2)/ \ep+ ...$, $\lambda_{3}^{-\ep} 
\rightarrow 1-\ep\,{\rm ln}(\lambda_3)+ ...$, we will find unregulated singularities as $\lambda_3 
\rightarrow 0$.  The most convenient method for separating the overlapping singularities in $\lambda_2$ 
and $\lambda_3$ is sector decomposition~\cite{Binoth:2000ps,Heinrich:2002rc,Binoth:2003ak}.  To illustrate this technique, a simple example 
suffices.  We consider the integral
\begin{equation}
\label{secex1}
I=\int_{0}^{1}{\rm d}x \,{\rm d}y \,x^{-1-\ep}y^{-1-\ep}
\left(x+y\right)^{-\ep}.
\end{equation}
The $1/x$ and $1/y$ factors cannot be expanded in plus distributions, as the logarithms from the expansion of $x+y$ 
will produce singular terms.  We split this integral into two parts,
\begin{equation}
\label{transformations}
I_1=\int_{0}^{1}{\rm d}x \int_{0}^{x}{\rm d}y  
\,x^{-1-\ep}y^{-1-\ep}\left(x+y\right)^{-\ep}, \,\,\,\,\,\,
I_2=\int_{0}^{1}{\rm d}y 
\int_{0}^{y}{\rm d}x  \,x^{-1-\ep}y^{-1-\ep}\left(x+y\right)^{-\ep}.
\end{equation}
In $I_1$ we set $y^{'}=y/x$, and in $I_2$ we set $x^{'}=x/y$.  Performing these variable changes, we find 
\begin{equation}
I_1=\int_{0}^{1}{\rm d}x \,{\rm d}y \,x^{-1-3\ep}y^{-1-\ep}
\left(1+y\right)^{-\ep}, \,\,\,\,\,\,
I_2=\int_{0}^{1}{\rm d}x \,{\rm d}y \,y^{-1-3\ep}x^{-1-\ep}
\left(1+x\right)^{-\ep}.
\end{equation}
The singularities in $x$ and $y$ are now separated in each integral, and can be extracted using Eq.~\ref{decomp}.  

One great advantage of this technique is the ease with which it can be automated.  The rules to determine 
when a term requires sector decomposition are simple; if the expression becomes singular when 
two (or more) 
variables $x,y \rightarrow 0$, but remains finite when either 
$x \rightarrow 0$ or $y \rightarrow 0$, 
then the transformations discussed below Eq.~\ref{transformations} 
should be performed.  Another advantage 
of sector decomposition is that it can be applied to fractional powers in addition to denominators, 
as illustrated in the example above.

We now discuss the application of this method to the denominator in 
Eq.~\ref{denomexam}, to show 
how it works in practice.  It is convenient to first separate the two singularities that can 
occur if $x \rightarrow 0$ or $x \rightarrow 1$ by splitting the integration as 
\begin{equation}
\label{pssplit}
\int_{0}^{1} {\rm d}x \rightarrow \int_{0}^{1/2} {\rm d}x + \int_{1/2}^{1} {\rm d}x \,\, ,
\end{equation}
and changing $x \rightarrow x^{'}$ in the second integration so that $x=1$ is mapped to $x^{'}=0$.  
Doing so for the three variables $\lambda_1$, $\lambda_2$, and $\lambda_3$ produces eight sectors.  
We focus on the sector where originally $\lambda_1,\lambda_2,\lambda_3 < 1/2$.  The denominator has the form 
\begin{equation}
{\cal D} = 
\frac{1}{\lambda_{1}^{2}\lambda_{2}\left[\lambda_2+\lambda_3(1-\frac{\lambda_1}{2})
	(1-\frac{\lambda_2}{2})\right]}.
\end{equation}
We now perform a sector decomposition in the variables $\lambda_2$ and $\lambda_3$, using the 
transformations given below equation Eq.~\ref{transformations}: in sector $a$ we set 
$\lambda_2 \rightarrow \lambda_2  \lambda_3$, and in sector $b$ we set 
$\lambda_3 \rightarrow \lambda_3  \lambda_2$.  Combining the Jacobian of the 
variable change with the denominator, we find the following expressions in each sector:
\begin{equation}
{\cal D}^{a}=\frac{1}{\lambda_{1}^{2}\lambda_2\lambda_3\left[\lambda_2+(1-\frac{\lambda_1}{2})
	(1-\frac{\lambda_2\lambda_3}{2})\right]} \,\,\,\,\, , \,
{\cal D}^{b}=\frac{1}{\lambda_{1}^{2}\lambda_2\left[1+\lambda_3(1-\frac{\lambda_1}{2})
        (1-\frac{\lambda_2}{2})\right]} .
\end{equation}
The terms in brackets are now finite in all limits; the denominators can be combined with the 
phase space measure, and the standard decomposition in plus distributions can be used to extract 
singularities.  Note that the above transformations must also be performed in the integration 
measure.  After splitting the integration as in Eq.~\ref{pssplit}, the measure contains 
terms of the form $(1-\lambda_i/2)$.  After sector decomposition, these become $(1-\lambda_i\lambda_j/2)$.  
If we had not performed this split, we would have produced terms of the form 
$(1-\lambda_i\lambda_j)$.  These are potentially singular when $\lambda_i,\lambda_j \rightarrow 1$, 
and would require further sector decomposition.

We must discuss 
two subtleties that can occur when using the method presented above.
The representation of ${\cal I}_{4}$ we have derived is 
convenient for expressions that do 
not contain $s_{13}$ or $s_{14}$ (see Eq.~\ref{invmasses}) in the denominator.  
In such terms, it is difficult to extract singularities 
in $\lambda_2$, $\lambda_3$, $\lambda_4$ that 
appear after integrating over $\lambda_5$.  One can 
always re-map the momenta of the final state particles
in a given diagram 
in such a way that $s_{14}$ never appears in the denominator.
For those terms that contain $s_{13}$, 
we first  bring the limits of 
the $s_{13}$ integration from 0 to 1 using the transformation
\begin{equation}
\label{nonlin}
{\hat \lambda}_5=\frac{s_{13}-s_{13}^{-}}{s_{13}^{+}-s_{13}^{-}}\, \frac{s_{13}^{+}}{s_{13}}.
\end{equation}
We then derive the following expression for the four particle phase space:
\begin{eqnarray}
\label{I4rep2}
\hat{{\cal I}}_{4}&=& {\cal N}_{4}\int_{0}^{1} {\rm d}\lambda_1{\rm d}\lambda_2{\rm d}\lambda_3{\rm d}\lambda_4{\rm d}{\hat \lambda}_5 \,
	\delta(\lambda_1-\lambda_{1}^{'})\delta(\lambda_2-\lambda_{2}^{'})\delta(\lambda_3-\lambda_{3}^{'})
	\delta(\lambda_4-\lambda_{4}^{'})\delta({\hat \lambda}_5-\hat{\lambda_{5}}^{'})  \nonumber \\ & & 
	\times \,
	\left[\lambda_1\left(1-\lambda_1\right)\left(1-\lambda_2\right)\right]^{1-2e} 
	\left[\lambda_2\lambda_3\left(1-\lambda_3\right)\lambda_4\left(1-\lambda_4\right)\right]^{-\ep}
	\left[{\hat \lambda}_5
\left(1-{\hat \lambda}_5\right)\right]^{-1/2-\ep} \nonumber \\ & & \times \, s_{13}({\hat \lambda}_5)
	\left[s_{13}^{+}s_{13}^{-}\right]^{-1/2-\ep}
\left\{{(1-\hat \lambda}_5)
\left(s_{13}^{+}-s_{13}^{-}\right)+s_{13}^{-}\right\}^{2\ep}.
\end{eqnarray}
Factors of $s_{13}$ that appear in the denominator are cancelled by the 
Jacobian of the non-linear transformation of Eq.~\ref{nonlin}, and 
the remaining ${\hat \lambda_5}$ integration does not produce dangerous 
singularities. Therefore, sector decomposition  of 
the ${\hat \lambda_5}$ integration is never required. 

One further complication exists.  Using the expressions in Eq.~\ref{s13lims}, we find
\begin{equation}
\label{manifold}
\left[s_{13}^{+}s_{13}^{-}\right]^{-1/2-\ep}
=\left(1-\lambda_1\right)^{-1-2\ep}|\lambda_2\left(1-\lambda_3\right)
	\left(1-\lambda_4\right)-\lambda_3\lambda_4|^{-1-2\ep}.
\end{equation}  
This expression is singular on a manifold of points in the interior of the phase space.  We wish to move these 
singularities to the boundary of the integration region.  To do so, we first note that the singularity occurs when 
\begin{equation}
\label{singpoint}
\lambda_4 \rightarrow \lambda_{4}^{s}=\frac{\lambda_2\left(1-\lambda_3\right)}{\lambda_3+\lambda_2
\left(1-\lambda_3\right)};
\end{equation}
this value is always in the integration region.  We can therefore split the $\lambda_4$ integration into two parts,
\begin{equation}
\int_{0}^{1} {\rm d}\lambda_4 = \int_{0}^{\lambda_{4}^{s}} {\rm d}\lambda_{4} + \int_{\lambda_{4}^{s}}^{1} 
{\rm d}\lambda_4,
\end{equation}
and then bring the integration limits back to 0 and 1.  Doing so produces two integrals, 
$\hat{{\cal I}}^{a}_{4}$ and $\hat{{\cal I}}^{b}_{4}$, with all singularities moved to the boundaries 
of the integration regions; these can be extracted using the sector decomposition technique 
discussed above.

To summarize, we derive analytic results for the double-real radiation corrections by following these steps: 
\begin{itemize}
\item We derive a factorized parametrization of the $1 \to 4$ phase-space as 
in Eq.~\ref{I4rep1} or Eq.~\ref{I4rep2}, in terms of kinematic 
variables which range from $0$ to $1$. 
\item We remove singularities from inside the allowed phase-space region to the boundaries by splitting appropriately the integrations and mapping them back to the $[0,1]$ interval. 
\item We then apply sector decomposition to disentangle the overlapping singularities. 
\item Finally, we extract the $\ep$ poles in terms of plus distributions. 
\end{itemize}

\section{The unitarity check}
Having discussed the four particle phase space and the technique of 
sector decomposition, we now illustrate our method by considering two examples 
of double real emission integrals with four propagators.  These are the most 
complicated phase space integrals that appear in $1 \rightarrow 4$ processes.  
We check our calculation of the double real emission corrections using their 
contributions to the imaginary parts of three-loop propagator diagrams.  From the optical theorem we know that the imaginary parts of such diagrams are given by the sum 
of all cuts, where all possible combinations of internal propagators are put on-shell.  
The required cuts also include real-virtual and virtual-virtual ones.  These are 
simple to compute, as are the imaginary parts of the propagator diagrams.  
We can therefore derive analytic expressions for the real-real cuts, which 
we can compare with the results we obtain using the methods presented in the 
previous Sections.  The checks we perform in this section involve inclusive 
integrations over the real emission phase space, in order to compare with the
imaginary part of the relevant propagator diagrams.  We will demonstrate in 
the next Section that the implementation of jet functions into our method is 
simple in both principle and practice.

We begin by considering the maximally planar propagator 
integrals which arise from Feynman graphs as the one shown in 
Fig.~1. 
The sum of all cuts for these diagrams contributes to the $e^+e^-$ 
total cross-section. 
\begin{figure}[htb]
\begin{center}
\begin{picture}(0,40)(0,0)
  \Oval(0,0)(20,40)(0)
\Photon(-50,0)(-40,0){1.8}{3}
\Photon(40,0)(50,0){1.8}{3}
\Gluon(-13,19)(-13,-20){1.8}{6}
\Gluon(13,19)(13,-20){1.8}{6}
\end{picture}
\end{center}
\caption{A planar propagator-type diagram contributing to the $e^+e^-$ 
cross-section at NNLO.}
\end{figure}
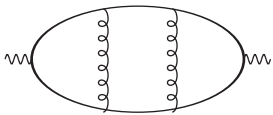 
Here, we examine the underlying scalar integral of these diagrams after 
we set their numerator to one.  
The analytic expression for this integral can be found using 
MINCER~\cite{Gorishnii:gt}; 
it is finite, and therefore its imaginary part vanishes.  The sum of all 
possible cuts of this diagram must also vanish.  This diagram has 
three distinct virtual-virtual cuts: the interference of a two-loop planar vertex correction 
to $\gamma^* \rightarrow q \bar{q}$ with the tree level contribution diagram together with 
its complex conjugate, and the square of the one-loop vertex correction to this process.  
It has four real-virtual cuts: two copies of the tree-level contribution to 
$\gamma^* \rightarrow q \bar{q} g$ interfered with the vertex correction to this process 
with the gluon radiated off the opposite quark leg, and two copies of its complex conjugate.  
Finally, it has two real-real cuts: the tree-level contribution to $\gamma^* \rightarrow 
q \bar{q} gg$ with both gluons radiated from a single quark line interfered with the diagram 
where both gluons are radiated from the opposite quark line, together with the complex conjugate 
of this contribution. 

The first virtual-virtual term, the two-loop vertex correction and its 
complex conjugate, is \cite{gonsalves}
\begin{equation}
V_p = \left ( \frac{\Gamma(1+\ep)}{(4\pi)^{d/2}} \right )^2 
{\cal R}_2
\left ( 
\frac{1}{2\ep^4} - \frac{4\zeta_2}{\ep^2} + 
\frac{10 \zeta_3}{\ep} - \frac{38\zeta_2^2}{5}
\right ),
\end{equation}
where ${\cal R}_2$ is the the two-particle phase space introduced 
in Section II.  The second virtual-virtual contribution, the square of the one-loop vertex 
correction, is
\begin{equation}
V_p^{1l} = 
\left ( \frac{\Gamma(1+\ep)}{(4\pi)^{d/2}} \right )^2 
{\cal R}_2
\left ( 
\frac{1}{\ep^4} - \frac{2\zeta_2}{\ep^2} - 
\frac{4 \zeta_3}{\ep} - \frac{4\zeta_2^2}{5}
\right ).
\end{equation}
The four real-virtual contributions give
\begin{equation}
RV_p = \left ( \frac{\Gamma(1+\ep)}{(4\pi)^{d/2}} \right )^2 
{\cal R}_2
\left (
\frac{-2}{\ep^4} + \frac{8\zeta_2}{\ep^2} - 
\frac{20 \zeta_3}{\ep} - \frac{156\zeta_2^2}{5}
\right ).
\end{equation}

The sum of the above contributions with a minus sign 
should equal the sum of the real-real cuts of this diagram.  We obtain
\begin{eqnarray}
R_p = && \left ( \frac{\Gamma(1+\ep)}{(4\pi)^{d/2}} \right )^2 
{\cal R}_2
\left (
\frac{1}{2\ep^4} - \frac{2\zeta_2}{\ep^2}
+\frac{14\zeta_3}{\ep} + \frac{198\zeta_{2}^{2}}{5}
\right ) 
\nonumber \\
&& =  \left ( \frac{\Gamma(1+\ep)}{(4\pi)^{d/2}} \right )^2 
{\cal R}_2 \left (
\frac{0.5}{\ep^4}-\frac{3.2899}{\ep^2}
+\frac{16.829}{\ep}+107.15.
\right ).
\end{eqnarray}
In terms of the invariant masses introduced in the previous 
Section, this contribution should be equal to 
\begin{equation}
R^{\rm num}_p = 
2 \left \langle \frac{1}{s_{13} s_{134} s_{24} s_{243}} \right \rangle,
\end{equation}
where the factor of two indicates the sum of both real-real cuts. This integral involves 
the invariant mass $s_{13}$; its calculation therefore requires 
extensive use of sector decomposition and the reparameterization 
$\lambda_5 \to \hat \lambda_5$ discussed in the previous 
Section.  Using these techniques, and numerically integrating the result 
using VEGAS \cite{Lepage:1980dq}, we obtain
\begin{eqnarray}
R^{\rm num}_{p} =\left ( \frac{\Gamma(1+\ep)}{(4\pi)^{d/2}} \right )^2 
{\cal R}_2 & &
\left (
\frac{0.5}{\ep^4}
+\frac{\left(-0.8\pm 9.2\right)\cdot 10^{-5}}{\ep^3}
-\frac{3.2909\pm 0.0018}{\ep^2}
+\frac{16.827\pm 0.010}{\ep} \right. \nonumber \\ & &
+107.12\pm 0.07 \left.
\right ).
\end{eqnarray}
We have included the VEGAS errors for those terms which require non-trivial integrations.
The agreement between the analytic and the numerical 
results is better than $0.1\%$ for all terms considered, and the differences are consistent 
with the integration errors.

We now consider the non-planar topologies shown in Fig.~2. 
Once again we focus on the scalar integral which is obtained from these 
topologies by setting their numerator to unity.
The imaginary part of this 
integral vanishes; therefore, we can again 
verify our calculation of the double real emission contribution using 
sector decomposition by checking the cancellation of all possible cuts.  This diagram has two 
virtual-virtual cuts: the interference of the two-loop non-planar vertex correction 
with the tree level $\gamma^* \rightarrow q\bar{q}$ diagram, and its complex conjugate.  It 
has four real-virtual cuts, each of 
which involves the one-loop box correction to 
$\gamma^* \rightarrow q\bar{q}g$ interfered with the tree-level contribution.  Finally, 
it has five real-real cuts, of three distinct types: two cuts involving the emission 
of two gluons off a single quark line interfered with the emission of two gluons off the 
opposite quark line; two cuts where a radiated gluon splits into a $q\bar{q}$ pair; 
one cut where a gluon is radiated from each quark line.  Since we are considering 
only scalar diagrams, the first four cuts give identical answers.
\begin{figure}[htb]
\begin{center}
\begin{picture}(0,40)(0,0)
  \Oval(0,0)(20,40)(0)
\Photon(-50,0)(-40,0){1.8}{3}
\Photon(40,0)(50,0){1.8}{3}
\Gluon(-13,19)(14,-20){1.8}{6}
\Gluon(14,19)(-13,-20){1.8}{6}
\end{picture}
\end{center}
\caption{A non-planar propagator-type diagram contributing to 
the $e^+e^-$ cross-section at NNLO.}
\end{figure}
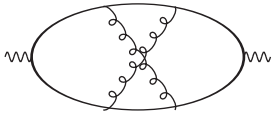 

The virtual-virtual cuts sum to \cite{gonsalves}
\begin{equation}
V_{np} = 
\left ( \frac{\Gamma(1+\ep)}{(4\pi)^{d/2}} \right )^2 
{\cal R}_2 
\left (
\frac{2}{\ep^4}-\frac{38\zeta_2}{\ep^2}
-\frac{54\zeta_3}{\ep} +\frac{966\zeta_2^2}{5}
\right ).
\end{equation}
The real-virtual contribution involves the integration of 
the one-loop box diagram with one leg off-shell.  There 
are two distinct ways of computing this diagram: either 
analytically using e.g. a Mellin-Barnes representation, 
or numerically using sector decomposition in a fashion identical 
to our approach to the real-real terms.  The second method is 
particularly convenient, as it allows restrictions on the final-state 
phase space to be imposed easily.  We now illustrate this technique.

The expression for the one-loop scalar box diagram
with one external leg off-shell, valid to all orders in $\ep$, is
\begin{eqnarray}
B && = \frac{2}{\ep^2 (4\pi)^{d/2}}
\frac{\Gamma(1-\ep)^2 \Gamma(1+\ep)}{\Gamma(1-2\ep)}
\frac{1}{st} 
\left [ 
 (-t)^{-\ep} F_{21}(1,-\ep,1-\ep,-\frac{u}{s})
\right. \nonumber \\ && \left. 
+ (-s)^{-\ep} F_{21}(1,-\ep,1-\ep,-\frac{u}{t})
- (-M^2)^{-\ep} F_{21}(1,-\ep,1-\ep,-\frac{M^2 u}{st})
\right ].
\label{eqbox}
\end{eqnarray}
Here, $M^2$ is the virtuality of the off-shell leg, and 
$s,t,u$ are the usual Mandelstam variables. We set $M^2=1$ in what follows and choose
the Mandelstam variables to be $s = s_{q \bar q} = \lambda_2$, 
$t = s_{\bar q g} = (1-\lambda_2)\lambda_1$,
$u = s_{ \bar q g} = (1-\lambda_2) (1-\lambda_1)$, where we have used our notation from Section II. 
The expression in Eq.~\ref{eqbox} 
must be integrated over the three particle phase, 
together with an additional propagator $1/s_{\bar q g}$ coming from the interference 
with the tree-level diagram, and the three-particle phase space in 
Eq.~\ref{phase3}.  
It is convenient to proceed in the following way. Using the integral representation 
for the hypergeometric function,
\begin{equation}
F_{21}(1,-\ep,1-\ep,z) =
\frac{\Gamma(1-\ep)}{\Gamma(-\ep)}
\int \limits_{0}^{1} {\rm d} t
\frac{ t^{-\ep-1}}{1-tz},
\end{equation}
we can write the required integrals over the phase space and 
the auxiliary variable $t$ in a form that is directly amenable 
to sector decomposition. It is then a simple task to numerically 
compute the expansion of the real-virtual corrections in powers 
of $\ep$ using the techniques described above. The analytic result for the 
real-virtual cut was derived for the purpose of checking our calculation 
based on sector decomposition; summing the four cuts, we obtain
\begin{equation}
RV_{np} = 
\left ( \frac{\Gamma(1+\ep)}{(4\pi)^{d/2}} \right )^2 
{\cal R}_2 
\left (
-\frac{10}{\ep^4}+\frac{128\zeta_2}{\ep^2}
+\frac{356\zeta_3}{\ep}-\frac{1108\zeta_2^2}{5}
\right ).
\end{equation}

Requiring the cancellation of all cuts, we derive the following analytic expression
for the real-real contribution:
\begin{eqnarray}
 R_{np} && = 
\left ( \frac{\Gamma(1+\ep)}{(4\pi)^{d/2}} \right )^2 
{\cal R}_2 
\left (
\frac{8}{\ep^4}-\frac{90\zeta_2}{\ep^2}
-\frac{302\zeta_3}{\ep}+\frac{142\zeta_{2}^{2}}{5}
\right )
\nonumber \\
&& = \left ( \frac{\Gamma(1+\ep)}{(4\pi)^{d/2}} \right )^2 
{\cal R}_2 
\left (
\frac{8}{\ep^4}-\frac{148.04}{\ep^2}
-\frac{363.02}{\ep}+76.84
\right ).
\end{eqnarray}
This result should be equal to the following sum of real-real integrals:
\begin{equation}
R^{\rm num}_{np} = 
 \left \langle \frac{1}{s_{13} s_{12} s_{24} s_{34}} \right \rangle
+
4 \left \langle \frac{1}{s_{13} s_{134} s_{23} s_{234}} \right \rangle.
\end{equation}
Computing the above integrals  using sector decomposition, we obtain
\begin{eqnarray}
R = 
\left ( \frac{\Gamma(1+\ep)}{(4\pi)^{d/2}} \right )^2 
{\cal R}_2 & &
\left (
\frac{8}{\ep^4}-
\frac{\left(5.8\pm 6.4\right)\cdot 10^{-4}}{\ep^3}
-\frac{148.06\pm 0.03}{\ep^2}-\frac{363.03\pm 0.07}{\ep} \right. \nonumber \\ & &+ 77.1\pm 0.4
\left. \right ).
\end{eqnarray}
The result found using sector decomposition is again consistent with that found 
by demanding the cancellation of all cuts, although the numerical precision 
of the finite piece is slightly worse.  This can be improved with 
a more sophisticated numerical integration technique.

We therefore conclude that our method can accomodate the most 
difficult real-real phase space integrals needed for $1 \rightarrow 4$ processes.  
We next consider in detail the $N_f$ dependent contributions to the 
$e^+e^- \rightarrow$ 2, 3, and 4 jet cross sections.  This example addresses the 
two remaining issues we must confront to fully validate our method: that jet 
functions can be implemented simply, and that bookkeeping of the 
sectors ({\it i.e.}, the expressions for the $s_{ij}$ in each sector as 
a function of the rescaled $\lambda_i$) can be peformed.

\section{$N_f$ dependent contribution to  $e^+e^- \to 2,3,4~{\rm jets}$}

In this Section we illustrate our method in a realistic NNLO example.  
We compute the $N_f$ dependent 
contributions to the 2 jet cross section at NNLO.  
When we wish to compute jet cross sections, we must include in our matrix 
elements a jet function, denoted by $F_J$ in Section II, that determines 
whether a given configuration contains 2, 3, or 4 jets.  This function 
takes the invariant masses of all partonic pairs as its arguments.  After 
splitting our result for the double real emission contribution into 
sectors, the invariant masses take different forms in terms of the $\lambda_i$ 
in each sector.  This presents bookkeeping issues that must be addressed.  
We prove here that we can handle this problem by considering a realistic 
example.  Taken in conjunction with the calculation in the previous Section 
of the most difficult integrals that appear in $e^+e^- \rightarrow$ 2 jets, 
the reader should be convinced of the power of our approach.

An example of the diagrams that contribute to the $N_f$ dependent terms of 
$e^+e^- \rightarrow$ jets at ${\cal O}(\alpha_{S}^{2})$ is shown in 
Fig.~3.  This diagram, together with the remaining contributions where 
the internal bubble is attached to a single quark line, contains both 
virtual-virtual and real-real cuts.  At ${\cal O}(\alpha_{S}^{2})$ we 
must also consider the coupling constant renormalization of the 
${\cal O}(\alpha_{S})$ result.  To present numerical results we must 
also choose a jet algorithm; we use the JADE algorithm, with a 
separation parameter $y=0.1$.
\begin{figure}[htb]
\begin{center}
\begin{picture}(0,40)(0,0)
  \Oval(0,0)(20,40)(0)
\Photon(-50,0)(-40,0){1.8}{3}
\Photon(40,0)(50,0){1.8}{3}
\Gluon(0,9)(0,20){1.8}{2}
\Gluon(0,-9)(0,-20){1.8}{2}
  \Oval(0,0)(10,10)(0)
\end{picture}
\end{center}
\caption{An example of an $N_f$ dependent diagram at NNLO.}
\end{figure}
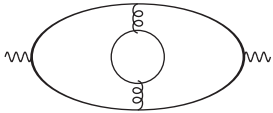 

The virtual correction only contributes to the two-jet configuration. Therefore, we write
\begin{equation}
\frac{\sigma_V}{\sigma_0} = \delta_{j,2}
N_f \left (\frac{\alpha_s}{\pi}\right )^2
\left (
\frac{1}{18\ep^3}
+\frac{11}{54 \ep^2}
+\frac{1}{\ep} \left (- \frac{11}{18}\zeta_2
+\frac{269}{324} \right )
+ \frac{5423}{1944}-\frac{13}{27}\zeta_3
-\frac{121}{54}\zeta_2
\right ),
\end{equation}
where the Kronecker delta indicates the restriction to the 2 jet cross section.  
We compute the double real emission using the approach described in the previous sections, 
and extract the singularities prior to integration over any kinematic variables.
Implementing the jet algorithm and performing the integrations over the five-dimensional 
phase-space numerically,  we obtain
\begin{eqnarray}
\frac{\sigma_R}{\sigma_0}  = 
N_f \left (\frac{\alpha_s}{\pi}\right )^2
& & \left ( 
\delta_{j,2} 
\left [ -\frac{\left(5.5553\pm 0.0005\right)\cdot 10^{-2}}{\ep^3}
-\frac{0.20369\pm 0.00005}{\ep^2}
+\frac{0.4180\pm 0.0005}{\ep} \right.\right. \nonumber \\ & & 
+ 4.808\pm 0.003
\left. \right ] 
-\delta_{j,3} \left ( 
\frac{0.41005\pm 0.00016}{\ep}
+ 2.9377\pm 0.0018 \right ) \nonumber \\ & &
+ \left(1.4561\pm  0.0018\right)\cdot 10^{-3}
\delta_{j,4}
\left. \right ).
\end{eqnarray}

The ${\cal O}(\alpha_s)$ cross-section combines 
the virtual correction and the single real emission.  We need this contribution 
to ${\cal O}(\ep)$ to derive its contribution to the NNLO cross section. 
Using the results in Section II, 
we derive
\begin{eqnarray}
\frac{\sigma^{(1)}}{\sigma_0}  =
\left (\frac{\alpha_s}{\pi}\right )
N_f & & \left (
\left ( -1.4597\pm 0.0013 -\left(9.242\pm 0.004\right)~\ep \right )\delta_{j,2} 
+ \right. \nonumber \\ & & \left ( 2.4575\pm 0.0012 + \left(6.115\pm 0.003\right)~\ep   \right )
\delta_{j, 3}
\left. \right ).
\end{eqnarray}
The ${\cal O}(\alpha_s^2)$ contribution to the $e^+e^-$ annihilation 
into hadrons is then written as
\begin{equation}
\sigma^{(2)} = \sigma_V + \sigma_R 
-\frac{\beta_0}{\ep}\left(\frac{\alpha_s}{\pi} \right) \sigma^{(1)}.
\end{equation}
The last term comes from the renormalization of the strong 
coupling constant in the ${\cal O}(\alpha_s)$ cross section; 
we need only keep the $-N_f/6$ term in the 
beta-function.  Adding these contributions, we obtain
\begin{eqnarray}
 \frac{\sigma^{(2)}}{\sigma_0} = 
N_f & & \left (\frac{\alpha_s}{\pi}\right )^2 
\left [ 
\delta_{j,2} 
\left (
\frac{\left(2.6\pm 4.6\right) \cdot 10^{-6}}{\ep^3}
+\frac{\left(1.4\pm 5.5\right) \cdot 10^{-5}}{\ep^2}
- \frac{\left(3.1\pm 5.2\right)\cdot 10^{-4}}{\ep} \right.\right. \nonumber \\ 
& & + 1.799\pm 0.003
\left. \right )
+\delta_{j,3} 
\left ( \frac{\left(-0.5\pm 2.6\right) \cdot 10^{-4}}{\ep} -1.917\pm 0.017 \right )
\nonumber \\ & & + \left(1.456\pm 0.002\right) \cdot 10^{-3}\delta_{j,4}
\left. \right ].
\end{eqnarray}
As we see, the divergences associated with various pieces 
disappear, with small remnants consistent with the integration errors. 
The cancellation occurs independently for the 2 and 3 jet cross-sections, as required. Finally, 
adding together the 2, 3, and 4 jet 
cross sections, we obtain the total hadronic cross section
\begin{equation}
\frac{\sigma^{(2)}}{\sigma_0}  
= \left(-0.117\pm 0.003\right)~N_f \left (\frac{\alpha_s}{\pi}\right )^2,
\end{equation}
which agrees with the known analytic result~\cite{book}:
\begin{equation}
\frac{\sigma^{(2)}}{\sigma_0}  
= \left (\frac{2}{3}\zeta_3 - \frac{11}{12}
\right )~N_f \left (\frac{\alpha_s}{\pi}\right )^2
 = -0.115~N_f \left (\frac{\alpha_s}{\pi}\right )^2.
\end{equation}
Again, the integration error of the finite piece can be improved with 
a more sophisticated numerical integration technique.  We conclude that our 
method can be applied succesfully to compute differential quantities.

\section{Conclusions}

We have presented a new technique for computing double real emission corrections at NNLO.  
Our method uses sector decomposition of the four particle phase 
space, together with an expansion in plus distributions, to extract the phase-space singularities without any analytic 
integrations, and preserves the exact kinematics of the partonic 
event.  The expressions for the matrix elements obtained with this 
approach can be used as building blocks 
for Monte Carlo event generators.

A phenomenologically attractive feature of our method is  that 
constraints on the final-state phase space, including various 
jet algorithms, can be implemented simply. This makes it
possible to study radiative corrections to  quantities of 
direct experimental relevance. 
The method is completely automated, and flexible.  
It can be applied  to any QCD or electroweak process with  
massless particles in the final state, where  the singularities 
from double-real unresolved radiation must be extracted.

We have illustrated our approach using $e^+e^- \rightarrow$ jets at ${\cal O}(\alpha_{S}^{2})$ 
as an example.  We have considered the most complicated phase space 
integrals that appear.  We have explicitly checked our results for those 
integrals by performing an inclusive numerical integration over 
the phase-space and comparing with analyting results obtained by using 
the unitarity method.  We have also demonstrated that our method is capable of calculating differential 
quantities  at NNLO by  deriving the $N_f$ dependent contributions 
to the ${\cal O}(\alpha_{S}^{2})$ cross section for 
$e^+e^- \rightarrow$ jets; this includes $e^+e^- \rightarrow$ 2 jets at NNLO, 
$e^+e^- \rightarrow$ 3 jets at NLO, and $e^+e^- \rightarrow$ 4 jets at LO.  
The $1/\ep$ poles were cancelled numerically, and the finite piece for the 
inclusive cross section agrees with results in the literature.

Results for the non-$N_f$ contributions will be given elsewhere.  
We have already presented here the calculation of the most difficult 
contributions needed for these terms. In addition, the bookkeeping of the various 
sector decompositions, and the numerical integrations, have already been addressed here.

While a direct application of our formalism to more complicated phase-spaces
is a  viable option, one could also use it profitably in conjunction with 
a dipole formalism. Sector decomposition can be applied to the 
dipoles to extract the $1/\ep$ singularities they contain, without the need for an analytic integration. The remainder 
can be integrated numerically. Sector decomposition of the finite terms should also improve the numerical stability of 
the dipole approach. 

There are several possibilities 
to develop the method further. It is interesting to investigate
its direct application to $1 \to 5$ processes.
It is also important for many applications to study the 
factorization of the phase space when  
massive particles appear in the final state. 
Although the parameterization 
of the phase space is certainly more complicated in those cases, we do 
not anticipate any significant limitations of the method.  We also expect
that the number of required sector decompositions 
will be reduced in the presence of massive particles.

Our method is a promising new technique for computing real radiation contributions to NNLO cross sections, and allows to obtain phenomenological results 
vital for the future  of precision high energy physics.  
We look forward to the  application of our method to many important 
collider physics processes.

\medskip
\noindent
{\bf Acknowledgments}: 
The work of C. A. is supported by the DOE under grant number 
DE-AC03-76SF0515.  The  work of K. M. is partially supported by the 
DOE under grant number  DE-FG03-94ER-40833
and  the Outstanding Junior Investigator Award DE-FG03-94ER40833.
  The work of F. P. is supported by NSF grants 
P420D3620414350 and P420D3620434350.  F. P. thanks the University 
of Hawaii at Manoa 
for kind hospitality during the completion of this work.

\medskip
\noindent
{\bf Note Added}: While this paper was being completed, a new
paper appeared which discusses the application of sector decomposition 
to inclusive phase space integrals~\cite{Gehrmann-DeRidder:2003bm}.


\begin{thebibliography}{10}

\bibitem{glover}
For a review, see T.~Gehrmann, hep-ph/0310178,  and references therein.

\bibitem{Kosower:2002su}
D.~A.~Kosower,
Phys.\ Rev.\ D {\bf 67}, 116003 (2003)
[arXiv:hep-ph/0212097].


\bibitem{Kosower:2003cz}
D.~A.~Kosower,
Phys.\ Rev.\ Lett.\  {\bf 91}, 061602 (2003)
[arXiv:hep-ph/0301069].


\bibitem{Weinzierl:2003fx}
S.~Weinzierl,
JHEP {\bf 0303}, 062 (2003)
[arXiv:hep-ph/0302180].

\bibitem{Weinzierl:2003ra}
S.~Weinzierl,
JHEP {\bf 0307}, 052 (2003)
[arXiv:hep-ph/0306248].

\bibitem{Ellis:1980wv}
R.~K.~Ellis, D.~A.~Ross and A.~E.~Terrano,
Nucl.\ Phys.\ B {\bf 178}, 421 (1981).


\bibitem{Giele:1991vf}
W.~T.~Giele and E.~W.~N.~Glover,
Phys.\ Rev.\ D {\bf 46}, 1980 (1992).


\bibitem{Fabricius:1980fg}
K.~Fabricius, I.~Schmitt, G.~Schierholz and G.~Kramer,
Phys.\ Lett.\ B {\bf 97}, 431 (1980).



\bibitem{Gutbrod:1983qa}
F.~Gutbrod, G.~Kramer and G.~Schierholz,
Z.\ Phys.\ C {\bf 21}, 235 (1984).




\bibitem{Catani:1996vz}
S.~Catani and M.~H.~Seymour,
Nucl.\ Phys.\ B {\bf 485}, 291 (1997)
[Erratum-ibid.\ B {\bf 510}, 503 (1997)]
[arXiv:hep-ph/9605323].


\bibitem{Binoth:2000ps}
T.~Binoth and G.~Heinrich,
Nucl.\ Phys.\ B {\bf 585}, 741 (2000)
[arXiv:hep-ph/0004013].

\bibitem{Heinrich:2002rc}
G.~Heinrich,
Nucl.\ Phys.\ Proc.\ Suppl.\  {\bf 116}, 368 (2003)
[arXiv:hep-ph/0211144].


\bibitem{Binoth:2003ak}
T.~Binoth and G.~Heinrich,
arXiv:hep-ph/0305234.



\bibitem{book} R.K.~Ellis, W.J.~Stirling and  B.R.~Webber, 
{\em QCD and Collider Physics}, Cambridge University Press, 1996.



\bibitem{Gorishnii:gt}
S.~G.~Gorishnii, S.~A.~Larin, L.~R.~Surguladze and F.~V.~Tkachov,
Comput.\ Phys.\ Commun.\  {\bf 55}, 381 (1989).


\bibitem{gonsalves}
R.~J.~Gonsalves, Phys. Rev. {\bf D28}, 1542 (1983).

\bibitem{Lepage:1980dq}
G.~P.~Lepage,
``Vegas: An Adaptive Multidimensional Integration Program,''
preprint CLNS-80/447.


\bibitem{Gehrmann-DeRidder:2003bm}
A.~Gehrmann-De Ridder, T.~Gehrmann and G.~Heinrich,
arXiv:hep-ph/0311276.


\end{thebibliography}

\end{document}